\documentclass[prd,showpacs,twocolumn]{revtex4}
\usepackage{amsfonts}
\usepackage{amssymb}
\usepackage{amsmath}
\usepackage[all]{xy}
\usepackage{graphicx}
\usepackage{amssymb}

\newcommand{\be}{\begin{equation}}
\newcommand{\ee}{\end{equation}}
\newcommand{\bn}{\begin{eqnarray}}
\newcommand{\en}{\end{eqnarray}}

\newcommand{\dslash}{\partial\!\!\!/}

\newcommand{\ed}{\end{document}}

\usepackage{amssymb}

\begin{document}

\title{Searching for the dual of the Maxwell-Chern-Simons model\\
minimally coupled to dynamical U(1) charged matter}

\author{M. S. Guimar\~aes$^2$}
\author{T. Mariz$^1$}
\author{R. Menezes$^1$}
\author{C. Wotzasek$^2$}

\affiliation{$^1$Departamento de F\' \i sica, Universidade Federal da
Para\'\i ba\\
58051-970 Jo\~ao Pessoa, Para\'\i ba, Brasil\\
$^2$Instituto de F\'\i sica, Universidade
Federal do Rio de Janeiro\\
21945-970, Rio de Janeiro, Brazil}

\begin{abstract}
The possibility of dual equivalence between the self-dual and the Maxwell-Chern-Simons
(MCS) models when the latter is coupled to dynamical, U(1) fermionic charged matter is examined. The proper coupling in the self-dual model
is then disclosed using the iterative gauge
embedding approach.
We found that the self-dual potential needs to couple directly to the Chern-Kernel of the source in order to establish this equivalence besides the need for a self-interaction
term to render the matter sector unchanged.

\end{abstract}

\maketitle

This paper deals with the concept of duality in three dimensional models with Chern-Simons term,  coupled to dynamical matter.
The study of such models have provided deep insights in unrelated
areas as particle physics and condensed matter, both from the theoretical
and phenomenological points of view \cite{DJT}.
Duality, on the other hand, is an useful concept in field theory and statistical mechanics since there are very few analytic tools available for studying non-perturbative properties of systems with many degrees of freedom.
One can easily make some elementary observations that hint towards the importance of the duality in D=3.
In this regard we mention the high temperature asymptotic of four dimensional
field theory models and the understanding of the universal behavior of the
Hall conductance in interacting electron systems. In particular this result
has been of great significance in order to extend the bosonization program
from two to three dimensions with important phenomenological
consequences \cite{boson}.

Certain theories, among them gauge theories admit description
in terms of different sets of potentials, the relation between these sets being called duality transformation.
This transformation typically maps solitons to fundamental fields and can therefore translate a non-perturbative problem to a solvable perturbative one.
In D=4 Maxwell theory there are two different such descriptions in terms of distinct potential one-form but the theories are identical. 
Dual theories may, however, not be identical. The duality between D=3 Maxwell and free scalar field theory being a well known example. Still, they are both described by second-order actions. 
Another well studied example in D=3 is the duality between Maxwell-Chern-Simons model (MCS) and the the Self-dual model (SD) \cite{TPvN} which is the subject of this investigation. However, what is more stringent in this case is the fact that while the gauge invariant MCS model is a second-order theory for the potentials, the SD model is described by a first-order theory, albeit for the field components. Such a feature, we will see, has striking consequences for the dual map when the models are coupled to external sources and fields. It is worth of observation that to study this duality also in the presence of sources is very important because physical observables are only obtained through measurements that critically depend on the couplings.

This paper is devoted to study issues of duality when the Maxwell-Chern-Simons model (MCS) is minimally coupled to dynamical fermionic sources. Similar questions have been tackled before by considering the Self-Dual model (SD) minimally coupled to dynamical fermionic matter \cite{GMdS} and also in \cite{Anacleto:2001rp} for bosonic matter as well.
This result is well illustrated by the following duality diagram,
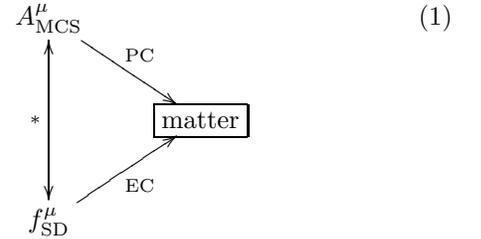
\be
 \label{eq1}
\xy \xymatrix{
                 A^\mu_{\rm{MCS}} \ar[dr]^{\rm PC} \ar[dd]_{*} & \\
          						  & *+[F-]{\rm{matter}}\\
                f^\mu_{\rm{SD}}	\ar[ur]_{\rm EC} \ar[uu] &	\\}				 
          						 \endxy 
\ee
that shows the self-dual field $f^\mu_{\rm{SD}}$ coupled electrically (EC) with the matter fields while the  MCS field $A^\mu_{\rm{MCS}}$ has a Pauli-type coupling (PC), as shown in \cite{GMdS, Anacleto:2001rp}.

These issues have been critically reviewed in \cite{Dalmazi:2003sz} who observe that due to the lack of gauge symmetry in the SD model, more general couplings should be allowed. However, the search for duality transformation when the gauge invariant MCS model is minimally coupled to the dynamical matter has remained an open question. This situation, illustrated in the following diagram,
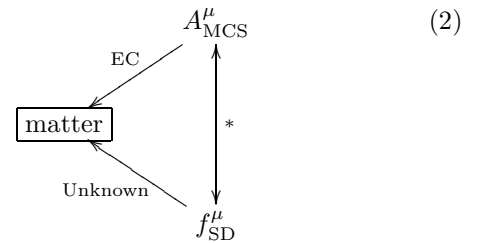
\be \label{eq2}
\xy \xymatrix{
                & A^\mu_{\rm{MCS}} \ar[dl]_{\rm EC} \ar[dd]^{*}  \\
          						   *+[F-]{ \rm{matter}}\\
                & f^\mu_{\rm{SD}}	\ar[ul]^{\rm Unknown} \ar[uu] 	\\}				 
          						 \endxy 
\ee
should therefore disclose the {\it unknown} coupling to the SD fields leading to electrical (or minimal) coupling in the gauge invariant side of the duality.
Besides its intrinsic interest in order to establish the correct coupling in the SD side of the duality,
this study is also important in order to define the functional generator which is meaningful for the measurement question discussed above.  Moreover, in order to complete the full program of duality with sources initiated in \cite{GMdS} it is mandatory to disclose such couplings.

The difficulty in the resolution of this problem is as follows.  When the matter currents are minimally coupled to fields in a first-order theory \cite{GMdS, Anacleto:2001rp}, such as the SD model, they are mapped, through duality, to their derivatives, $J^\mu \to \epsilon^{\mu\nu\lambda}\partial_\nu J_\lambda$, which, together with the presence of Thirring like term, guarantees the invariance of the matter dynamical content. The coupling induced in the MCS model then becomes the well known Pauli-term and represents an electric dipole interaction. However, when the opposite situation is considered, matter current minimally coupled to the second-order side of the duality, the problem seems to have a more complex status.
In fact, the SD field is expected to couple to a sort of {\it inverse derivative operator} of the current \footnote{This operator is in fact the well known Hopf operator and has been brought in the context of quantum Hall effect in F. Wilczek and A. Zee, Phys.Rev.Lett.51 (1983) 2250.}. Such an object, known as the Chern-Kernel of the source, although well defined mathematically as,
\be
\label{ckd}
J_\mu = \epsilon_{\mu\nu\rho} \partial^\nu \omega^\rho \, ,
\ee
lacks a significant physical meaning. With this definition for the Chern-Kernel $\omega_\mu$, the current is automatically conserved but the Chern-Kernel is ambiguous.
Indeed, the kernel transformation as
\be
\label{ckt}
\omega_\mu \to \omega_\mu + \partial_\mu \chi
\ee
leaves the current invariant. Geometrically the Chern-Kernel is seem as the world-sheet of the Dirac string attached to the charge.  As so, the property (\ref{ckt}) is directly related to the unobservability of the Dirac string.

However, the kernel symmetry (\ref{ckt}) seems to bring another unexpected problem.  Because of it the coupled SD model could acquire the status of a gauge theory. We will see that a related feature in gauge theories is the responsible for the solution of this difficulty.  In fact the MCS model needs gauge fixing.  On the other hand, to solve (\ref{ckd}) for the Chern-Kernel we need to give sense for the inverse of the operator $\epsilon_{\mu\nu\rho} \partial^\nu $ that is undefined because of the presence of a zero-mode. 
A possible regularization of the symmetry (\ref{ckt}) is the following,
\be
\label{regularization}
\omega_\mu = \left[ \frac{1}{\epsilon^{\mu\nu\rho} \partial_\nu}\right]_{reg} J^\rho \equiv \frac{\epsilon_{\mu\nu\rho} \partial^\nu}{\Box} J^\rho
\ee
that automatically satisfy,
\be
\partial_\mu \, \omega^\mu = 0
\ee
this way eliminating the ambiguity in the Chern-Kernel.

It is interesting at this juncture to relate the ambiguity problem of the Chern-Kernel $\omega_\mu$ in terms of the current $J_\mu$, just mentioned, with the gauge symmetry displayed by the MCS model.  This situation is in fact deeply rooted to the existence of a duality between these two models.
To study duality in this context, different techniques have been developed since its original postulation \cite{DJ}. Recently we have proposed the {\it gauge embedding} approach to deal with this question in the presence of sources \cite{IW}. This approach is also interesting since it naturally discloses a factorization for the propagator of the MCS model in terms of the propagators for the SD model and a pure Chern-Simons model.
This is a worthy way to understand the meaning of this equivalence since the MCS theory gives origin to second-order differential equations while the SD model is a first order theory. There is therefore an extra solution in the first that is lacking in the second.  Therefore, in a sense that will soon become clear, the extra solution in the MCS must be trivial. This is indeed the meaning of the gauge symmetry present in the former but not in the later.

In order to put the comments above in solid grounds, let us then quickly review the gauge embedding approach to duality, in the free case \cite{IW}. This also serve a second purpose as a review of the technique. To this end, let us write the SD model as
\begin{equation}
\label{PB11}
{\cal L}_{SD} = \frac 12\, f^\mu D^{SD}_{\mu\nu} f^\nu \, ,
\end{equation}
where
\begin{equation}
\label{PB13}
D_{\mu\nu}^{SD}=\left[{\cal R}^{SD}_{\mu\nu}\right]^{-1}=m^{2}\eta_{\mu\nu} - m\epsilon_{\mu\nu\lambda}\partial^\lambda
\end{equation}
is the inverse propagator for the SD model. From the Euler vector of the SD model\footnote{The Euler vectors $K_\mu$, are defined by the independent variations of
the action, whose kernel gives the equations of motion.}
\begin{equation}
\label{PB12}
K_\mu = D^{SD}_{\mu\nu} f^\nu =\left( m^{2}\eta_{\mu\nu} - m\epsilon_{\mu\nu\lambda}\partial^\lambda\right) f^\nu
\end{equation}
we obtain the equations of motion as the kernel of the Euler vector, $K_\mu = 0$.

The approach of \cite{IW,Anacleto:2001rp} works by iteratively inducing the required invariance into the original model through the remotion of the obstruction to gauge symmetry which, after the elimination of some auxiliary fields, gives the dual of the original non-invariant model as
\begin{eqnarray}
\label{PB14}
 ^*{\cal L} &=& {\cal L}_{SD} - \frac{1}{2\, m^2} K_\mu K^\mu\nonumber\\
&=& \frac 12 A^\mu D_{\mu\nu} A^\nu - \frac 1{2m^2} A^\mu D^{2}_{\mu\nu}A^\nu \nonumber\\
&=& \frac 1{2m} A_\mu\left[(\epsilon^{\mu\rho\lambda}\partial_\lambda)D_{\rho}^\nu\right]A_\nu \nonumber\\
&=&{\cal L}_{MCS}
\end{eqnarray}
where we have relabeled $f_\mu \to A_\mu$ to reflect the embed gauge character of the new variable and called $D_{\mu\nu}^{SD} = D_{\mu\nu}$ to simplify the notation.
From here we observe that the $D_{\mu\nu} A^\nu=0$, a solution of the SD model, is a solution for the MCS model as well. However there is another solution in the form
\begin{equation}
\label{PB15}
\epsilon_{\mu\nu\lambda}\partial^\nu A^\lambda = 0,
\end{equation}
which is pure gauge, that is not present in (\ref{PB11}).

The propagator ${\cal R}_{\mu\nu}^{MCS}$ for the MCS model has therefore been factorized as,
\bn
{\cal R}_{\mu\nu}^{MCS} &=& \left[(\epsilon^{\mu\rho\lambda}\partial_\lambda)D_{\rho}^\nu\right]^{-1}\nonumber\\
&=& \left[\frac 1{\epsilon^{\mu\rho\lambda}\partial_\lambda}\right] \, {\cal R}_{\rho\nu}^{SD}\, .
\en
The propagating degrees of freedom of both theories, described by ${\cal R}_{\mu\nu}^{SD}$ in (\ref{PB13}), clearly coincide but the MCS has a pure gauge freedom that is not manifest in the SD, a well known fact. What is of importance here is that the gauge freedom manifests itself in the MCS model through the pure Chern-Simons component of the propagator which is the same as the one manifest by the Chern-Kernel above. As mentioned, gauge-fixing of the MCS model automatically regularizes the zero-mode for the pure Chern-Simons operator and vice-versa,
\bn
\left[{\cal R}_{\mu\nu}^{MCS}\right]_{g.f.} 
= \left[\frac 1{\epsilon^{\mu\rho\lambda}\partial_\lambda}\right]_{reg} \, {\cal R}_{\rho\nu}^{SD}\, .
\en

We are now ready to consider the problem posed above.  Our strategy will be as follows. Starting with an ansatz action, representing the SD model coupled to the Chern-Kernel of the electric current, we apply the gauge embedding program to obtain the MCS model with minimal coupling. This will complete the duality picture initiated in \cite{Anacleto:2001rp}.

Let us then consider the following ansatz for the Chern-Kernel coupling for the SD
\begin{equation}
\label{PB30}
{\cal L}^{(e)}_{SD} = \frac {m^2}2 \,{\left( f_\mu - \frac em \,\omega_\mu \right)}^2 + \frac m2 f^\mu \epsilon_{\mu\lambda\nu}\partial^\lambda f^\nu + {\cal L}_D ,
\end{equation}
where
\begin{equation}
\label{PB21}
{\cal L}_{D} =  \bar{\psi}(i\partial\!\!\! /  -M)\psi \; ,
\end{equation}
describes the free Dirac field. 
The (regularized) Chern-Kernel $\omega_\mu$ is given in terms of the fermionic fields as
\be
\omega_\mu =\frac{\epsilon_{\mu\nu\rho} \partial^\nu}{\Box} \,\bar\psi \gamma^\rho \psi
\ee
Following the embedding approach we compute the Euler vector   
\begin{equation}
\label{PB31}
 K_\mu = m^2 \left( f_\mu - \frac em \,\omega_\mu \right) +  m \,\epsilon_{\mu\lambda\nu}\partial^\lambda f^\nu  
\end{equation}
and write the dual model as (after the relabel $f_\mu\to A_\mu$)
\begin{eqnarray}
\label{PB32}
& & ^*{\cal L}^{(e)} = {\cal L}^{(e)}_{SD} - \frac 1{2m^2} K_\mu^2 \nonumber\\
& &            =- \frac {1}4 F_{\mu\nu}^2 + \frac m2 A_\mu\epsilon^{\mu\lambda\nu}\partial_\lambda A_\nu + e\, A_\mu J^\mu + {\cal L}_D ,
\end{eqnarray}
which shows as claimed, the minimal coupling between the MCS-field $A_{\mu}$ and the fermionic source. It is noteworthy that this time the duality transformation did not induce any Thirring like current-current interaction. However, a similar feature has appeared, this time as a self-interacting term for the Chern-Kernel albeit in the SD model. Still, the matter dynamics remains unchanged as will be next verified. However, before that, it is important to consider this result in the perspective of previous contributions to the subject. In \cite{RZ}, Rey and Zee have discussed the self-duality of the MCS-Proca action including the contribution of vortices and magnetic monopoles. The modification $f^\mu \rightarrow f^\mu -(e/m)\omega^\mu$ used to consider the direct coupling with the Chern-Kernel has been used in \cite{RZ} to take account of topologically non-trivial field configurations -- vortices in 2+1 dimensions -- and a factorization of the action into its self and anti-self dual components was found. In contrast, the factorization found here display the self-dual component of the MCS model and a pure Chern-Simons part. The nonlocal operator used here to define the Chern-Kernel in terms of the fermionic fields has appeared before in \cite{EW} and \cite{CH} to discuss the $SL(2,Z)$ symmetry present in the MCS action. A discussion along the same lines is to be found in \cite{BD1} and \cite{BD2} together with a study of the particle-vortex duality with applications to quantum Hall effect.


To verify the invariance of the matter dynamics we start computing the fermionic field equations in the self-dual case. To this end let us rewrite (\ref{regularization}) as
\begin{eqnarray}
\omega^{\mu}(x) =  \int d^3y G(x-y)\varepsilon^{\mu\nu\rho}\partial^{(y)}_{\nu}J_{\rho}(y),
	\label{mcsad40}
\end{eqnarray} 
where $\Box_{(x)}G(x-y) = \delta(x-y)$. The equation of motion for the fermionic field is
\begin{eqnarray}
0 &=& \frac{\delta S_{AD}[\bar{\psi}]}{\delta \bar{\psi}(x)} = (i\gamma^{\mu}\partial_{\mu} - M)\psi(x) \nonumber\\
 &-& \int d^3y \left(ef^{\mu}(y) - e^2\omega^{\mu}(y)\right)\frac{\delta \omega_{\mu}(y)}{\delta \bar{\psi}(x)},
	\label{mcsad41}
\end{eqnarray} 
where $S_{SD}[\bar{\psi}] = \int d^3y {\cal L}^{(e)}_{SD}$. Taking the functional derivative of (\ref{mcsad40}) we obtain
\begin{eqnarray}  & & (i\dslash - M)\psi(x) =\nonumber\\
 & &  =\! me \!\!\!\int \!\!d^3\!y G(x\!-\!y)\varepsilon_{\mu\nu\rho}\partial^{\nu}\!\!\left[f^{\rho}(y)\! -\! \frac em\omega^{\rho}(y)\right]\!\!\gamma^{\mu}\psi(x). \nonumber
\end{eqnarray}
Next we get rid of the self-dual field, in favor of the fermion fields, from its equations of motion,
\begin{eqnarray}
  f_{\mu} = \frac em {\cal R}^{SD}_{\mu\nu}\omega^{\nu},
	\label{mcsad44}
\end{eqnarray} 
and use that
\begin{eqnarray}
    \frac 1{m} \varepsilon_{\nu\sigma\rho}\partial^{\sigma} = \eta_{\nu\rho} - \frac1{m^2}\left[{\cal R}^{SD}_{\nu\rho}\right]^{-1},
  \label{mcsad34}
\end{eqnarray}
to obtain 
\begin{eqnarray}  & & e\,\varepsilon_{\mu\nu\rho}\partial^{\nu}\left(f^{\rho} - \frac em \omega^{\rho}\right)=\nonumber\\  & &  = e^2(m \eta_{\mu\rho} - \frac1m{\cal R}^{-1}_{\mu\rho})\frac em{\cal R}^{\rho\sigma}\omega_{\sigma} - \frac{e^2}m J_{\mu}\nonumber\\  & &  = e^2{\cal R}_{\mu\sigma}\omega^{\sigma} - \frac{e^2}{m^2}\omega_{\mu} - \frac{e^2}{m}J_{\mu}, 	\label{mcsad45}
 \end{eqnarray}  
where ${\cal R}_{\mu\nu} ={\cal R}^{SD}_{\mu\nu}$. Going back to the symbolic matricial notation of (\ref{ckd}), the purely fermionic dynamics is given then as
\begin{eqnarray}
 && (i\gamma^{\mu}\partial_{\mu} - M)\psi = \nonumber\\
 &=& \frac {e^2}{\Box} \left[ m{\cal R}_{\mu\sigma}\omega^{\sigma} - \frac{\omega_{\mu}}m - J_{\mu}\right]\gamma^{\mu}\psi.
	\label{mcsad46}
\end{eqnarray}  

Let us consider next the fermionic dynamics of the MCS model described by (\ref{PB32}).  From there we obtain the matter equations of motion as,
\begin{eqnarray}
  (i\gamma^{\mu}\partial_{\mu} - M)\psi(x) = eA_{\mu}\gamma^{\mu}\psi(x).
	\label{mcsad48}
\end{eqnarray}
As before, to obtain the purely fermionic dynamics, we eliminate the gauge field $A_{\mu}$ by solving the gauge equations of motion,
\begin{eqnarray}
  \varepsilon^{\mu\nu\rho}\partial_{\nu}A_{\rho} = \frac em {\cal R}^{\mu\nu}J_{\nu},
	\label{mcsad49}
\end{eqnarray}
which, after gauge fixing ($\partial_{\mu}A^{\mu} = 0$) becomes,
\begin{eqnarray}
  A_{\mu} = \frac em \left[ -\frac{\varepsilon_{\mu\nu\rho}\partial^{\nu}}{\Box}\left({\cal R}^{\rho\sigma}J_{\sigma}\right)\right],
	\label{mcsad50}
\end{eqnarray}
Substitution in (\ref{mcsad48}) results in
\begin{eqnarray}
  &&(i\gamma^{\mu}\partial_{\mu} - M)\psi = \frac{e^2}{m} \left[ \frac{\varepsilon_{\mu\nu\rho}\partial^{\nu}}{\Box}\left({\cal R}^{\rho\sigma}J_{\sigma}\right)\right]\gamma^{\mu}\psi\nonumber\\
  & =& \frac{e^2}{\Box} \left[ m{\cal R}_{\mu\sigma}J^{\sigma}-\frac1{m^2} J_{\mu}\right]\gamma^{\mu}\psi\nonumber\\
  &=&\frac {e^2}{\Box} \left[ m{\cal R}_{\mu\sigma}\omega^{\sigma} - \frac{\omega_{\mu}}m - J_{\mu}\right]\gamma^{\mu}\psi \, ,
	\label{mcsad51}
\end{eqnarray}
which coincides with (\ref{mcsad46}). This result gives an, {\it a posteriori}, proof of the ansatz (\ref{PB30}) by leaving the fermionic dynamics unaffected. As anticipated, the matter behaves as expectator under duality.


In conclusion in this paper we resumed the study of the dual equivalence between the self-dual
model \cite{TPvN} and the Maxwell-Chern-Simons theory \cite{DJT} coupled
to dynamical fermionic matter, using the iterative gauge
embedding procedure \cite{IW}.
In the former studies \cite{GMdS,Anacleto:2001rp} where it was the SD model that appeared coupled minimally to dynamical matter, the dual mapping into the MCS theory showed that, (i) it exchanges the minimal coupling into a
non-minimal Pauli-type interaction and (ii) introduces a current-current
Thirring-like interaction to preserve the dynamics of the fermionic matter
sector.
In the present study we found that in order to have the gauge invariant MCS model minimally coupled to a conserved current the SD model needs to be coupled to the Chern-Kernel of the source.  Besides, a self-interacting kernel-kernel term also becomes necessary to preserve the dynamics of the fermionic sector.
The results of the the duality reported here are new and, in fact, quite surprising.
Although the presence of a coupling of the SD field with a sort of {\it inverse derivative} of the current could be anticipated, the necessity of a quadratic Chern-Kernel piece to preserve the matter dynamics was unexpected.
As far as we know, such a model and its properties have not been investigated before.

We finish with some worthy noticing observations.
In this study we focused on the case of minimal coupling of the source with the MCS, suggested by gauge invariance.
Such an investigation seems important in order that the observables of both theories could be compared. It is also important to obtain a better understanding of the connection between the {\it dual interactions}, like the dipole and minimal, from one side and the meaning of the direct Chern-Kernel interaction, on the other side. It is very important to mention that the completion the full duality program was dependent on this result. It is now possible to undertake a study of the quantization of the parameters involved in the models, e.g., charge and mass \cite{futuro}. Finally, it is worth mentioning the possibility of extending this program to other dimensions and to tensors of higher ranks completing the studies of \cite{Menezes:2002nj,Menezes:2003vz}.

\noindent ACKNOWLEDGMENTS: This work is partially supported by CNPq/PRONEX/FAPESQ, CAPES/PROCAD, FAPERJ and
FUJB, Brazilian Research Agencies.  The authors would like to thank D. Bazeia for many
suggestions.  CW thanks the Physics Department of UFPB the kind hospitality
during the course of this investigation.

\end{document}